\documentclass[article,onecolumn]{revtex4}
\usepackage{graphicx}
\newcommand\beq{\begin{equation}}
\newcommand\eeq{\end{equation}}
\newcommand\bear{\begin{eqnarray}}
\newcommand\eear{\end{eqnarray}}
\begin{document}

\title {External Electric Field Mediated Quantum Phase Transitions in 
One-Dimensional Charge Ordered Insulators: A DMRG Study} 
\author{Sudipta Dutta and Swapan K. Pati}

\address{Theoretical Sciences Unit and DST Unit on Nanoscience
\\Jawaharlal Nehru Centre For Advanced Scientific Research
\\Jakkur Campus, Bangalore 560064, India.}

\date{\today}
\widetext

\begin{abstract}
\parbox{6in}

{We perform density matrix renormalization group (DMRG) calculations 
extensively on one-dimensional chains with on site ($U$) as well as 
nearest neighbour ($V$) Coulomb repulsions. The calculations are carried 
out in full parameter space with explicit inclusion of the static bias 
and we compare the nature of spin density wave (SDW) and charge density 
wave (CDW) insulators under the influence of external electric field. 
We find that, although the SDW ($U>2V$) and CDW ($U<2V$) insulators 
enter into a conducting state after a certain threshold bias, CDW 
insulators require much higher bias than the SDW insulators for 
insulator-metal transition at zero temperature. We also find the 
CDW-SDW phase transition on application of external electric field. 
The bias required for the transitions in both cases decreases with 
increase in system size.}

\end{abstract}
\maketitle

\narrowtext

 Strongly correlated electronic systems in low-dimension are  
always very interesting because of their unique characteristics
like quantum fluctuations, lack of long range ordering etc
\cite{Geller,NI,Others,Wiegmann,Ogota,Schulz,Sumit}. The correlation 
effects in such low-dimensional systems lead to Mott or charge-ordered 
insulating states. The one-dimensional extended Hubbard 
model\cite{Hubbard,Vollhardt} with on-site Hubbard repulsion term, $U$, 
along with the nearest-neighbour Coulomb repulsion term, $V$, is a 
standard model which exhibits these different phases. This is perhaps 
the simplest possible model which can capture many interesting properties 
of strongly correlated systems. In the strong coupling limit, this model 
gives rise to two insulating phases, spin-density-wave (SDW) and 
charge-density-wave (CDW) which are separated by a first order phase 
transition line at $U\simeq2V$. On the other hand, in the weak coupling 
limit, the perturbative analysis shows that the transition is continuous
at $U=2V$\cite{Hirsch,Furusaki}. 

The effect of electric field on such insulating phases has attracted
much interest in recent times due to the practical applications in tuning 
their dielectric and piezoelectric properties\cite{Swapan_epl}. Many 
electronic conduction processes seem to suggest electric field induced 
phenomena, such as negative differential resistance in molecular 
electronics\cite{Tour,Lakshmi1}. Experiments on low-dimensional 
Mott-insulators with spin-density-waves and charge-ordered phases as 
ground states, suggest a collapse of insulating phase in presence of external 
electric field at finite temperatures
\cite{Taguchi,Yamanouchi,Tokura,CNR,Dong,Dicarlo,Monceau,Wu,Ong,Dumas,Gruner}. 
However, the breakdown in charge ordered phases is not due to 
Joule heating. Rather it has been argued that, it is the applied 
bias which generates a conduction pathway resulting in metallic 
characteristics\cite{CNR}. However, a tractable computational method 
which can take into account the static electric field and its response 
on the correlated extended electronic systems is still lacking.

In this article, we use the Density Matrix Renormalization Group 
(DMRG)\cite{White,Schollwock,SKPati} method which is known to be 
highly accurate for low-dimensional interacting systems with high 
precision. We have included electric field in the DMRG algorithm 
and have obtained ground and excited eigen states behaviors of 
one-dimensional CDW and SDW insulating systems with various system sizes. 
The static electric field is included as a ramp potential and we 
find that the electric field can induce an insulator-metal-insulator 
transition in both the cases, with a strong dependence on the 
Hamiltonian parameters. Instead of field per unit length, we consider
the total field, applied between two ends of a one-dimensional chain,
irrespective of the chain length. Similar consideration was applied
for SDW insulators, where we extrapolated
our finite size results to very large (effectively infinite) systems to
obtain thermodynamic behavior\cite{Sudipta}. In the present study on
insulating CDW phase, we find the CDW-SDW quantum phase transition 
along with the breakdown of insulating phase on application of bias.

We consider one dimensional strongly correlated chain described 
by the extended-Hubbard Hamiltonian,
\begin{eqnarray}
H &=& t\sum\limits_{i}(a^\dagger{_i}a_{i+1}+h.c) \nonumber \\
&+& U \sum\limits_{i}n_{i\uparrow}n_{i\downarrow} 
+ V \sum\limits_{i}(n_{i}-n_{av})(n_{i+1}-n_{av})
\end{eqnarray}

\noindent where t is the hopping term, $U$ is the onsite Hubbard 
repulsion term and $V$ is the nearest-neighbour Coulomb repulsion 
term. We set $t=1$ in all our calculations and express every 
energy in units of $t$. The $a^\dagger$ ($a$) is the creation 
(annihilation) operator and $n$ ($n_{av}$) is the number operator 
(avarage electron density on every site). The external electric field 
applied on the system has the form of a ramp potential, distributed 
over all the sites in such a way that the potential $F_{i}$ at site 
$i$ becomes $-\frac{F}{2}+i\frac{F}{N+1}$, where $F$ is the total
applied bias and $N$ is the total number of sites 
in the 1D chain. This form of the potential ensures that the 
bias varies between $-F/2$ to $F/2$ across the molecule. The 
potential adds an extra term $\sum\limits_{i}F_{i}a^\dagger{_i}a{_i}$ 
to the above Hamiltonian. In DMRG, we use density-matrix cut-off, 
$m = 140$. For $U=0$ and $V=0$, the problem can be exactly solved 
and we obtain the ground state and excitation spectrum in 
presence of bias using the tight-binding one-electron formalism. 
We perform our calculations keeping a fixed value of $U$ $(=5)$ and 
varying the $V$ from $0$ to $4$ with bias from $0$ to $10$ volts in 
steps of $0.5$ volts.

With increase in the value of $V$, one encounters three different 
regions, namely, spin-density wave $(2V<U)$, spin-density-wave to 
charge density wave crossover $(2V=U)$ and charge-density-wave 
$(2V>U)$. To understand the effect of electric field on the excited 
states of the system, for a given value of $U$ and $V$, we have 
computed the energies of the systems with one extra $(E(N+1))$ and 
one less electron $(E(N-1))$ than half-filling $(E(N))$ and have 
calculated the many body charge excitation gap as the difference 
between the energy required to add ($\mu_+$) and remove ($\mu_-$) 
electrons from the ground state\cite{insulator},

\begin{eqnarray}
\Delta_{charge}=\mu_{+}-\mu_{-}
\end{eqnarray}

\noindent where $\mu_{+}=E(N+1)-E(N)$ and $\mu_{-}=E(N)-E(N-1)$.
 
\begin{figure}
\centering
\includegraphics[scale=0.3, angle=270] {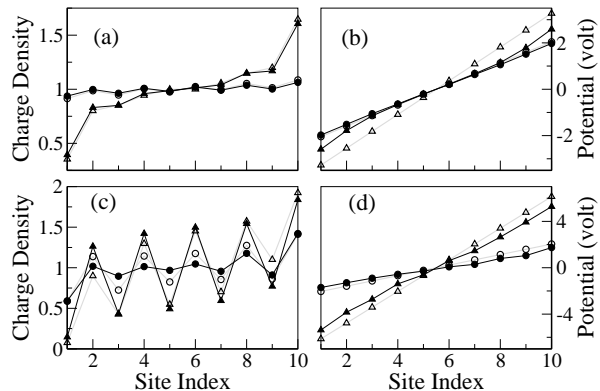}
\caption{The charge density $(a)$ and the potential $(b)$ on all the 
sites in SDW phase with $U=5$, $V=1$ and $F=5(circle)$ and 
$8(triangle)$. $(c)$ and $(d)$ correspond to the charge density and 
potential respectively in CDW phase with $U=5$, $V=3$ and $F=5(circle)$ 
and $15(triangle)$. Open and closed symbols correspond to calculations 
without and with self consistent Poisson's equation respectively for 
$N=10$.}
\end{figure}

To investigate the effect of polarization on the applied electric field, 
i.e., the screening of external electric field by the shifted electron 
density, we solve the self consistent Poisson's equation in SDW as well 
as in CDW phases. We have considered the boundary conditions so that the 
final electrostatic potential field extending from one electrode to 
another becomes different for different atomic sites and the atomic 
electron density will adjust to the field in such a way that the charges 
are stabilized locally. We start our self consistent calculation by 
assuming that the electrostatic field is a linear ramp function across 
the interface of electrode and the system. By solving the Schrodinger 
equation within many-body limit we obtain the charge density at every 
site and use that as an input in one-dimensional Poisson's equation. The 
diagonal elements of the Hamiltonian is then modified with the modified bias, 
obtained from Poisson's equation. Solving the modified Hamiltonian we obtain 
the charge densities and again use them as input in Poisson's equation. 
This is continued untill all the charge densities and all the site 
potential fields converge. In Fig.1 we plot both the on-site atomic charge 
densities and the spatial distribution of potentials for both SDW and CDW 
phases with explicit consideration of the screening of applied electric 
field and show the same, without considering the screening, for comparison. 
From Fig.1 it is clear that, in the presence of strong electronic 
correlations, the ramp nature of the external electric field is retained 
without any significant change of the charge density or potential profile 
with the consideration of polarization effect. Here we consider the system 
with $N=10$ and use exact diagonalization method and present the results 
for two different bias values. The ground state of the CDW phase is a
linear combination of two configurations (having alternating doubly 
occupied sites) with same weightage. So we take the average of the 
charge densities and the potentials of the two degenerate ground states
in case of CDW phase. Moreover, for CDW phase we consider fairly large
value of bias to investigate the breakdown region, which we discuss
later. To infer, large correlations reduces the Poisson's equation 
into Laplace equation with no significant variation of the charge 
density distribution which we observed in our earlier studies 
also\cite{Lakshmi2,Lakshmi3}. The inclusion of the effect of polarization 
can change the quantitative estimation of the results without any significant 
change of the physics. So, to avoid the computational expense of the
self consistent Poisson's equation calculations, we ignore the 
polarization effect in our further DMRG calculations with larger system 
sizes.

\begin{figure}
\centering
\includegraphics[scale=0.3, angle=270] {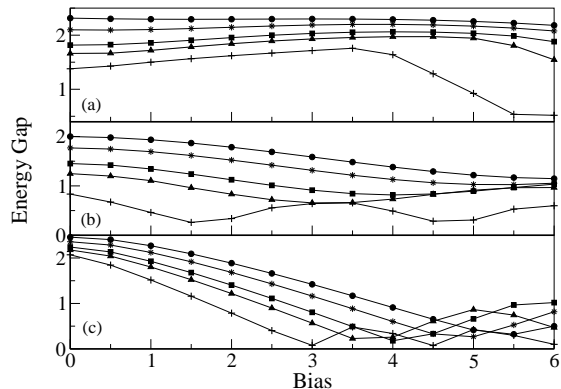}
\caption{Variation of charge gap as a function of bias for 
(a) CDW ($V=3$), (b) SDW-CDW crossover ($V=2.5$) and (c) SDW
($V=1$) insulators, for various systems sizes (N): $N=10 (circle), 
12 (star), 16 (square), 20 (triangle)$ and $40 (plus)$.}
\end{figure}

In Fig.2(a), (b), and (c) we have plotted the charge gap of 
1-D chains with different $N$ as a function of bias for CDW, 
SDW-CDW crossover region and SDW phases, respectively. For SDW
phase and SDW-CDW crossover region, the charge gap decreases
with increasing bias upto a minimum, whereas in case of 
CDW phase, the charge gap increases initially with increasing 
bias. In the latter case, the nearest-neighbour Coulomb
repulsion term, $V$, is higher than the previous two cases, and as 
a consequence, the electrons favour to remain paired in 
alternative sites than increasing its kinetic energy by hopping to the 
neighboring sites. The SDW phase has a charge gap of $~U$ at 
$U\rightarrow\infty$ limit. But the CDW phase with double occupancy 
at alternative sites for a half-filled 1-D chain has smaller charge gap 
than $U$ and at $U,V \rightarrow\infty$ limit, the charge gap is $~(U-V)$. 
Therefore, in the CDW phase, the $V$ is the stabilizing term and it can 
be proved unambiguously from the model Hamiltonian. Bias, however, tries 
to make the electrons hop in its direction and as a consequence 
the site occupancies start changing, leading to a higher 
charge gap. It is thus the hopping, which creates larger gap in 
charge-ordered phase inducing positive contribution from $V$. 
However, this initial increase is followed by a decrease in charge gap 
at quite a higher bias, although, the charge gap minimum can not be 
seen for all the system sizes in the bias window, we have considered. 
In case of SDW phase, the bias overcomes the effect of $U$ by forcing 
the electrons to hop in its direction and leads the system to a charge 
gap minima corressponding to a conducting state. So, it is clear from 
Fig.2 that, the CDW phase takes much higher bias to pass through 
a charge gap minimum than the SDW phase. The SDW-CDW crossover region 
shows combination of the properties of both SDW and CDW phases. It can 
also be seen from Fig.2 that, the increase in system size reduces the 
initial charge gap for all the three cases and reduces the bias 
corressponding to the first charge gap minimum in case of 
SDW phase and SDW-CDW crossover region.

\begin{figure}
\centering
\includegraphics[scale=0.3, angle=270] {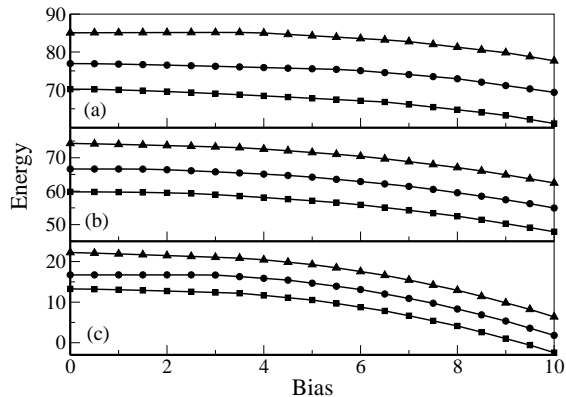}
\caption{Energies, $E(N+1)$ ($triangle$), $E(N)$ ($circle)$ and
$E(N-1)$ ($square$) as a function of bias for (a) $V=3$, 
(b) $V=2.5$ and (c) $V=1$, for the system with $40$ sites.}
\end{figure}

To understand the response of the ground state as well as the 
excited states under the influence of external electric field, 
we plot $E(N+1)$, $E(N)$ and $E(N-1)$ of a system with $40$ sites 
as a function of bias in Fig.3 for CDW phase ($V=3$), SDW-CDW 
crossover region ($V=2.5$) and for SDW phase ($V=1$). For all the 
three cases, the ground state energies do not change much upto a 
threshold bias. However, beyond that, the systems start to stabilize. 
The slope of the ground state energy is different from those of 
excited states for all the three phases and this difference in 
slope gives rise to collapse of insulating phase at some threshold 
bias, which is different for different phases and strongly depends 
on the system size. In contrast to the first two cases, although 
the CDW phase with one less electron starts getting stabilized 
with increasing bias, the system with one more electron destabilizes 
slightly in low bias followed by lowering in energy in higher bias 
values. It is due to the fact that, one more electron in CDW phase 
experiences more prominent elctronic repulsion for bias driven 
hopping because of higher $V$ value. As a consequence, the charge 
gap increases initially, however, beyond a certain bias value, 
required to nullify the effect of $V$, the gap reduces.

\begin{figure}
\centering
\includegraphics[scale=0.3, angle=270] {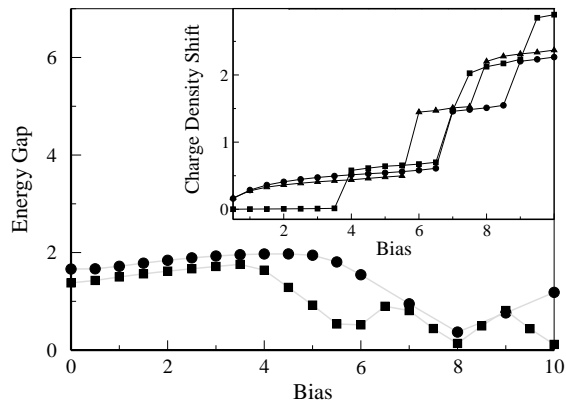}
\caption{Charge gap as a function of bias for system sizes, 
$N=20$ $(circle)$ and $N=40$ $(square)$ for $V=3$. Inset shows 
the average charge density shift per site as a function of bias for 
$N+1$ $(square)$, $N$ $(triangle)$ and $N-1$ $(circle)$
electrons for system size $N=40$.}
\end{figure}

For a clear understanding of the charge gap variation in CDW
insulating phase with increasing bias, in Fig.4 we plot the charge 
gap, $\Delta_{charge}$ as a function of bias for $V=3$, for finite 
systems with $N=20$ 
and $N=40$ sites. As can be seen clearly, the charge gap increases 
initially upto a certain bias and then it starts decreasing approching 
a minimum at some threshold bias. Further increase in bias drives the 
system to go through several minima. On the contrary, the charge gap 
in SDW phase starts decreasing with even small nonzero bias and it 
goes through several minima with increasing bias as evident from Fig.2 
as well as in our previous study\cite{Sudipta}. In CDW phase, the 
initial increase in charge gap is only due to the nearest neighbour 
Coulombic repulsion term which prevents the hopping of electrons. After 
a certain bias, as depicted in the Fig.4, the charge gap starts decreasing 
and then onwards the variation of charge gap with increasing bias looks 
very similar as found in the SDW phase. From this, we can anticipate 
qualitatively that, 
on application of external bias, the CDW phase undergoes a phase 
transition at some bias, $V_{trans}$, and beyond that starts to behave 
like a SDW phase. At higher bias values, the periodic oscillation of 
charge gap in charge ordered phase makes the system to go through several 
insulator-metal-insulator transitions. However, from Fig.4 it is clear 
that, the period of oscillation becomes narrower with increase in the 
size of the system as observed in our previous study\cite{Sudipta} on SDW 
phase.

\begin{figure}
\centering
\includegraphics[scale=0.3, angle=270] {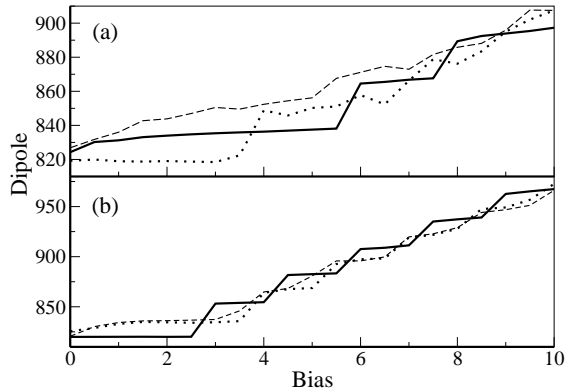}
\caption{The dipole moment as a function of bias for
(a) $V=3$ and (b) $V=1$ for $N+1(dotted$ $lines)$, $N(solid$ $lines)$ and 
$N-1(dashed$ $lines)$ electron states for a system size of $N=40$.}
\end{figure}

To understand the underlying reasons for such behaviour of the charge 
gap, in the inset of Fig.4, we plot average charge density shift per site 
for the $N=40$ one-dimensional chain with $N+1$, $N$ and $N-1$ electrons 
as a function of bias. As can be seen, 
although all the three states show staircase like behavior, for the system 
with one extra electron, the nature of the shift of the charge densities is 
completely different from the other two cases. The charge density shifts 
slowly at lower bias in case of half-filled system and the system with one 
less electron, whereas the charge density shift for the system with one extra 
electron is almost zero upto a certain bias, termed as $V_{trans}$, for this 
system, beyond which, the charge density shows a drastic shift. It is due to 
the fact that, the extra electron faces more prominent effect of Coulomb 
repulsion terms and hopping of electrons is possible only at some bias value 
which can overcome the electronic repulsion. After $V_{trans}$, the CDW phase 
again goes to an insulating phase, characterized by the plateau in the plot, 
which we anticipate to be a SDW phase. If we further increase the bias, after 
certain values, the shift in charge densities for all the three systems show 
sudden jump, which indicates kinetic stabilization, followed by plateaus, 
indicating reappearance of insulating phase. We argue that this repetetive 
period of charge gap with increase in bias is due to the charge stiffness. 
The external bias sweeps the charge densities towards one direction with 
nullification of repulsion terms upto the first minima. Beyond this, the bias 
tries to shift the charge densities further and the electronic repulsion 
terms again reappear and start opposing the kinetic stabilization, and as a 
consequence, the charge gap again increases and the system enters into 
an insulating phase. Thus, this repetetive metallic and insulating 
behaviour in different bias regions is a consequence of the interplay 
between the external bias and the electron-electron interaction terms. 
However, this oscillation of the charge gap with increase in bias is 
due to finite size effects. With increase in system size, the periodicity
narrows down and we can anticipate that, at infinite system size, all the 
minima will converge to the same bias value as evident from our previous 
study\cite{Sudipta}. We termed this bias value at thermodynamic limit as 
the critical bias, $V_{c}$. Note that, as we discussed earlier, we do not 
consider the effect of polarization on applied electric field in our 
calculations. It can only change the quantitative estimation of 
$V_{trans}$ and the critical bias, $V_c$ without any qualitative change 
in the basic physics.

To obtain a better insight, we plot the dipole moment of the ground state
and the states with one extra and one less electron for both SDW and 
CDW phases as a function of bias in Fig.5. It can be seen clearly that, 
the ground state dipole moment shows periodic jumps at the bias values, 
corresponding to the charge gap minima and the plateaus indicate the
successive insulating phases. So, whenever there is a shifting of charge 
densities i.e., sudden increase in the dipole moment, the system allows
the electrons to hop, leading to breakdown of insulating phase. This 
periodic jumps of the dipole moment is purely due to finite size effects. 
However, the CDW-SDW phase transition can not be understood from 
the ground state dipole moments.

\begin{figure}
\centering
\includegraphics[scale=0.3, angle=270] {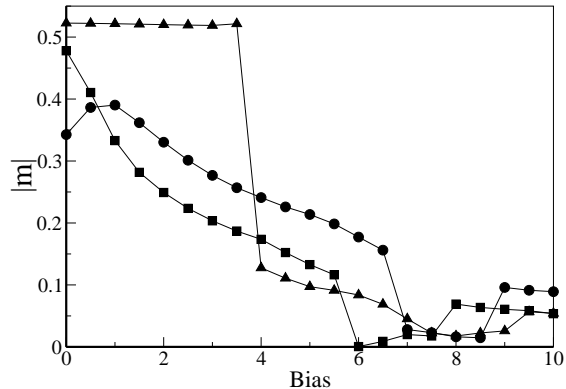}
\caption{Absolute value of the order parameter, $m$ for CDW phase ($V=3$) 
with $N+1$ ($triangle$), $N$ ($square$) and $N-1$ ($circle$) electrons as 
a function of bias.}
\end{figure}

To understand the CDW-SDW phase transition, we calculate the order 
parameter\cite{Hirsch} in terms of onsite charge densities 
($\langle n_{i} \rangle$) for a finite system, consisting of $40$ sites 
with half-filling and with one less and one more electron than half-filling.

\begin{eqnarray}
m &=& \frac{1}{N}\sum_{i}(-1)^{i} \langle n_{i} \rangle
\end{eqnarray}

\noindent and plot its absolute value in Fig.6. As can be seen, after 
a certain bias, the $|m|$ value for the system with one extra electron 
drops, indicating clearly a phase transition to SDW phase. The $|m|$ values 
for the other two cases also show drops at some bias values, which indicates 
the insulator to metal phase transition. From the above observations, we can 
conclude that, the system with one extra electron can directly indicate the 
CDW-SDW phase transition on application of external bias. Although the $|m|$ 
values show oscillations in Fig.6, we can interpret them as purely finite 
size effect, as described earlier. From these observations, we adequately 
conclude that, the extarpolation of our results to infinite chain length
will result into a single drop in the order parameter and take the CDW phase 
to a SDW phase, through the quantum phase transition at thermodynamic limit.

\begin{figure}
\centering
\includegraphics[scale=0.3, angle=270] {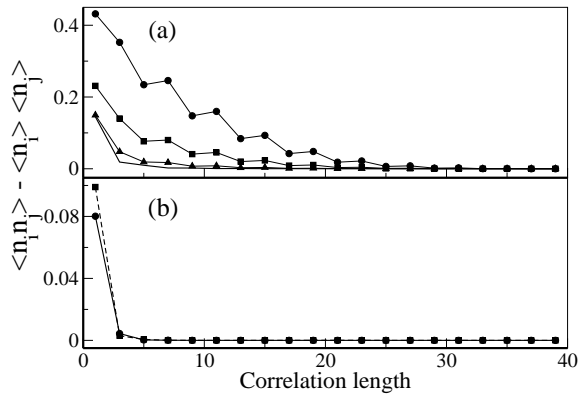}
\caption{The charge-charge correlation as a function of 
correlation length for (a) $V=3$ at bias values $0.0(circle)$, 
$0.5(square)$, $1.0(triangle)$ and $5.5(solid$ $line$ $without$ $symbol)$ 
volts and for (b) $V=1$ at bias values $0.0(circle)$ and $6.0(square)$ 
volts for the system with $40$ sites.}
\end{figure}

However, under the influence of external electric field which sweeps 
the electrons in its direction, the spatial charge distribution 
becomes nonuniform, resulting in mixture of locally confined domains 
of CDW and SDW orders. As a consequence, the spatial charge distribution 
becomes less significant to be considered as an order parameter. 
Instead, the charge-charge correlation function
($\langle n_{i}n_{j} \rangle - \langle n_{i} \rangle\langle n_{j} \rangle$)
becomes more informative to characterize the phases. In Fig.7, we have 
shown the ground state charge-charge correlation as a function of 
distance for both CDW and SDW phases for a system with $40$ sites. 
The nature of the zero-bias correlation is completely different in 
these two cases. In case of CDW phase, the correlation is known to be 
long range, whereas for SDW phase it is short ranged. Fig.7(a) clearly 
shows that, the correlation length for CDW phase decreases with the 
increase in bias and after a certain bias it behaves exactly like the 
SDW phase, depicted in Fig.7(b). This resemblence in the behavior of 
correlation function strongly indicates the CDW-SDW phase transition.

In conclusion, we have studied the application of external electric 
field on both the SDW and CDW insulating phases with on-site and 
nearest-neighbour Coulombic repulsion. We find that the external 
electric field can induce a insulator-metal transition in both the 
cases. Moreover, the electric field induces a CDW-SDW quantum phase 
transition as well. Increase in system size increases the sharpness 
of both the transitions. Our DMRG calculations allow us to study large 
systems with explicit application of electric field, with microscopic 
understanding of the insulator-metal transition in spin-density-wave 
and charge-density-wave insulators.

SD acknowledges the CSIR for the research fellowship and S. Lakshmi 
for illuminating discussions. SKP acknowledges the research support from 
CSIR and DST, Govt. of India.

\end{document}